\documentstyle[epsfig,cite]{mn}

\begin{document}

\title[Orbital, precessional and flaring variability of Cygnus X-1]
  {Orbital, precessional and flaring variability of Cygnus X-1}
\author[Brocksopp et al.]
  {C.~Brocksopp$^1$\thanks{email: cb@star.cpes.susx.ac.uk}, R.P.Fender$^2$, V.~Larionov$^3$, V.M.~Lyuty$^4$,  A.E.~Tarasov$^5$,
\newauthor 
G.G.~Pooley$^6$,W.S.~Paciesas$^7$, P.~Roche$^1$\\
$^1$Astronomy Centre, CPES, University of Sussex, Falmer, Brighton, 
BN1 9QJ\\
$^2$Astronomical Institute ``Anton Pannekoek'' and Center for High-Energy Astrophysics, University of Amsterdam, \\
\hspace{0.5cm}Kruislaan 403, 1098 SJ Amsterdam, The Netherlands\\
$^3$Astronomical Institute, St Petersburg University, 198904, St Petersburg, Russia\\ 
$^4$Sternberg Astronomical Institute, 119899, Moscow, Russia\\
$^5$Crimean Astrophysical Observatory, 334413, Nauchny, Crimea, Ukraine\\
$^6$Mullard Radio Astronomy Observatory, Cavendish Laboratory, Madingley Road, Cambridge CB3 0HE\\
$^7$NASA/Marshall Space Flight Center, Huntsville, AL 35812, USA\\}
\date{Accepted ??. Received ??}
\pagerange{\pageref{firstpage}--\pageref{lastpage}}
\pubyear{??}
\maketitle

\begin{abstract}

We present the results of a 2.5-year multiwavelength monitoring programme of Cygnus X-1, making use of hard and soft X-ray data, optical spectroscopy, {\it UBVJHK} photometry and radio data. In particular we confirm that the 5.6-day orbital period is apparent in all wavebands and note the existence of a wavelength-dependence to the modulation, in the sense that higher energies reach minimum first.  We also find a strong modulation at a period of $142 \pm 7$ days, which we suggest is due to precession and/or radiative warping of the accretion disc. Strong modulation of the hard and soft X-ray flux at this long period may not be compatible with simple models of an optically thin accretion flow and corona in the low state. We present the basic components required for more detailed future modelling of the system -- including a partially optically thick jet, quasi-continuous in the low state, the base of which acts as the Comptonising corona. In addition, we find that there are a number of flares which appear to be correlated in at least two wavebands and generally in more. We choose two of these flares to study in further detail and find that the hard and soft X-rays are well-correlated in the first and that the soft X-rays and radio are correlated in the second. In general, the optical and infrared show similar behaviour to each other but are not correlated with the X-rays or radio.
 
\end{abstract}


\section{Introduction}

Cygnus X-1 has been one of the most well-studied X-ray sources since its discovery in 1965 (Bowyer et al. 1965) and identification with the O9.7 Iab supergiant HDE226868 (Bolton 1972, Webster \& Murdin 1972). It has been particularly interesting as it provided the first observational evidence for the existence of black holes when it was found to be a spectroscopic binary with an orbital period of 5.6 days. The ephemeris was revised by Gies \& Bolton (1982) and the masses of the two components were later calculated as 17.8$M_{\sun}$ for the supergiant and 10.1$M_{\sun}$ for the black hole (Herrero et al. 1995). It is only recently that the ephemeris has been improved (Brocksopp et al. 1999, Sowers et al. 1998, LaSala et al. 1998), although this is still within the errors of the Gies \& Bolton (1982) result and does not alter the above mass estimates.

The supergiant is thought to emit a stellar wind as is typical for high mass stars. However, spectroscopic analysis has shown that the wind is attracted towards the black hole and so the focussed wind model of Friend \& Castor (1982) is appropriate (Gies \& Bolton 1986). In addition to this, many authors have found that the H$\alpha$ and He\,{\sc ii}\,$\lambda4686$ emission components originate at the L1 point and follow a 5.6-day orbital modulation out of phase with the photospheric lines of the supergiant (e.g. Aab 1983).

The radio counterpart was discovered by Braes \& Miley (1971). While at the time of discovery the radio flux was $\sim$20 mJy, it generally maintains a constant level of $\sim$14 mJy at cm wavelengths; a $\sim$3 mJy modulation at both the orbital period and sometimes at $\sim$150 days is superimposed on this (Pooley, Fender \& Brocksopp 1999). The radio emission shows a very flat, synchrotron  spectrum and may be produced via the conical jet model of Hjellming \& Johnston (1988).

In addition, the orbital period has been well-defined in the optical (Voloshina, Lyuty \& Tarasov 1997) and infrared (Leahy \& Ananth 1992, Nadzhip et al. 1996) and also in X-ray data collected by {\it Ariel V}/ASM (Holt et al. 1976), {\it RXTE}/ASM (Zhang, Robinson \& Cui 1996) and {\it CGRO}/BATSE (e.g. Paciesas et al. 1997).

Long-term variations have now been discovered in various X-ray binaries such as SS 433 (e.g. Abell \& Margon 1979), Her X-1 (Tannanbaum et al. 1972b) and LMC X-3 (Cowley et al. 1991) and interpreted as the precession period of the accretion disc in each system. It is still not understood exactly how this takes place, although there are a number of suggestions assuming very different mechanisms. Larwood (1998) uses a model invoking tidal forcing by the companion's gravitational pull on the solid disc; the ratio of observed orbital and precessional periods of a selection of long-period X-ray binaries appear to fit this model. However, Wijers \& Pringle (1998) argue that the models of e.g. Larwood (1998) do not fit the observations and do not explain how the whole disc can precess at the same rate, given that the precession frequency is a function of disc radius, or how the disc remains tilted. Instead, they follow the suggestions of Petterson (1977) and Iping \& Petterson (1990) that sufficient illumination from the centre could cause a disc to become to become warped and tilted. This radiative model can also account for the prograde precession seen in Cygnus X-2 which cannot be explained by tidal models (Wijers \& Pringle 1998, and references therein). 

A number of correlated events have been observed since the discovery of the radio counterpart. X-ray transitions from the soft to the low/hard state (see e.g. Esin et al. 1998 and references therein for definitions of the various states of X-ray binaries) occurred simultaneously with an increase in radio flux from $\sim$5 mJy to $\sim$20 mJy, levelling out at $\sim$15 mJy (Tananbaum et al. 1972a, Braes \& Miley 1976). This has been observed more recently by the VLA -- the radio flux increased from $\sim$5 mJy to $\sim$20 mJy simultaneously with the transition from the high/soft state back to the hard in August 1996 (Zhang et al. 1997b). Even more notable was the radio flare of 45 mJy that occurred in May 1975 (Hjellming, Gibson \& Owen 1975). This is the only time that the radio flux has been reported to increase so far above its mean level and, again, this was accompanied by a decrease in the soft X-rays. 

The state transitions have been explained in terms of changes in the inner disc radius and the advection models of Esin et al. (1998). During the usual hard X-ray state the inner disc radius is relatively large and hard X-ray emission is thought to be produced by the upscattering of disc photons by a Comptonised corona (Zhang et al. 1997a). During the high/soft state there is slightly increased mass flow, the corona becomes optically thin and the inner edge of the accretion disc extends to the last stable orbit surrounding the black hole. It is necessary that $\dot{M}$ increases only slightly since the bolometric X-ray luminosity is only $\sim10-20$\% higher during the high/soft state; however the increase in $\dot{M}$ may be more significant during the transition periods (when the hard X-rays are not quite at their minimum) as at this time the bolometric luminosity is signifcantly higher (Zhang et al. 1997a). It has also been suggested that the transition to the high/soft state is the result of a disc reversal -- theoretically possible for wind-driven systems such as Cyg X-1 (Zhang, Cui, Chen 1997). 

However, these models frequently fail to take the radio emission and orbital modulation of both radio and X-ray into account. Therefore, in this paper we present a modified version to account for the full spectrum.

Correlated events between various wavelength regimes have occurred in many X-ray binaries, including GRO J1655-40 (Tavani et al. 1996), Cygnus X-3 (McCollough et al. 1998) and possible hard X-ray/radio correlations have been found for GX339-4 (Hannikainen et al. 1998) -- a persistent black hole candidate X-ray binary very similar to Cyg X-1 in its X-ray and radio properties. In particular, GRS 1915+105 has become well-known for its radio plateau states which coincide with hard X-ray outbursts (Fender et al. 1999). It appears that during state changes in Cyg X-1 the radio and hard X-rays correlate whilst being anticorrelated with the soft X-rays. Therefore, full radio monitoring of Cyg X-1 through one of the transitions from low/hard state to soft and back again is necessary in order to see how the radio and X-ray behaviour compares with that of other systems. Clearly it is important to study these correlations if we are to gain knowledge of the emission mechanisms and hence obtain complete models for the systems.

\section{Observations}

\begin{figure*}
\vspace*{-1.5cm}
  \begin{center}
    \leavevmode  
    \psfig{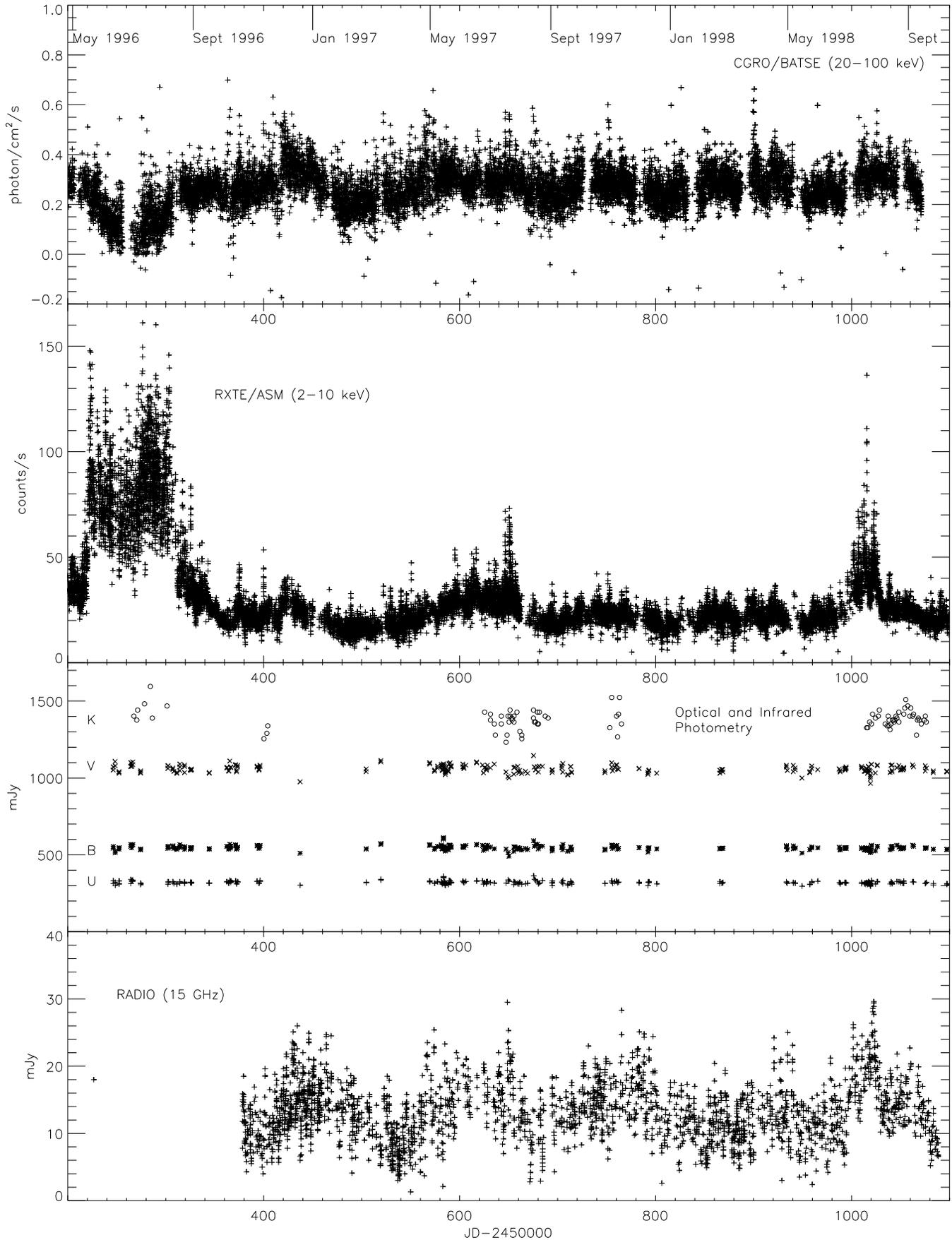}  
    \vspace{-0.5cm}  
    \caption{BATSE, ASM, optical, infrared and radio lightcurves for the full 2.5 years or our observations (MJD 50200--51100).}
    \label{fig:Lightcurves}
  \end{center}
\end{figure*}

\begin{table}
\caption{Summary of observations}
\label{Observations}
\begin{center}
\begin{tabular}{lcc}
\hline
Band&Source&Observation Period\\
\hline
Hard X-rays (20-100 keV)&{\it CGRO}& April 1996--Sept 1998\\
Soft X-rays (2-10 keV)&{\it RXTE}&April 1996--Sept 1998\\
{\it UBV}&SAI&June 1996--Sept 1998\\
Spectroscopy&CrAO&June 1996--Sept 1998\\
{\it JHK}&SPG+CrAO&June 1997-- Sept 1998\\
Radio&RT&Oct 1996--Sept 1998$^{*}$\\
\hline
\end{tabular}
\end{center}
CrAO -- Crimean Astrophysical Observatory\\
SAI -- Sternberg Astronomical Institute\\
SPG -- St. Petersburg University\\
RT -- Ryle Telescope, Cambridge, UK\\
$^{*}$ There is also one point from April 1996\\
\end{table}

Data were collected between April 1996 and September 1998, a summary of which can be found in Table 1.

To ease comparison between datasets the optical and infrared magnitudes have been converted into mJy using the O'Dell conversion factors (O'Dell et al. 1978).

\subsection{X-ray}
Our hard X-ray data came from the BATSE instrument on the Compton Gamma Ray Observatory ({\it CGRO}) and was processed using the standard BATSE earth occultation software. Further details can be found in Paciesas et al. (1997). Soft X-ray data has been obtained from the Rossi X-ray Timing Explorer ({\it RXTE}); we use the ASM public archive data from the web. A detailed description of the ASM, including calibration and reduction is published in Levine et al. (1996)

\subsection{Optical}
The photometric $UBV$ data were obtained by the pulse-counting photometer of the Crimean Laboratory (Nauchny, Crimea) of Sternberg Astronomical Institute; the 60 cm telescope was used. All observations were made with reference to the local standard BD $+35^{\circ}3816$, $V=9.976, B-V=0.590, U-B=0.064$ (Lyuty 1972).
 
Our spectra were taken at the Crimean Astrophysical Observatory using the coud\'{e} spectrograph of the 2.6 metre telescope. The detector was a CDS9000 (1024$\times$256 pixels) CCD array. All observations were made in the second order of a diffraction grating with reciprocal dispersion of 3\AA/mm and resolution of 25000. The typical exposure time for each spectrum totalled 40 minutes for H$\alpha$ and 1.5 hours for He\,{\sc ii}\,$\lambda 4686$ resulting in a S/N of $\sim$100. The spectra were reduced using standard flat field normalisation and sky subtraction techniques. Wavelength calibration was achieved using ThAr comparison spectra. The water lines were removed from the H$\alpha$ spectra and equivalent widths calculated.

\subsection{Infrared}
The infrared data were collected using an InSb photometer of St. Petersburg University which is attached to the 0.7m telescope of the Crimean Observatory. The detector operates at the temperature of liquid nitrogen. $\eta$ Cygni was used as the standard star and the measurements have an accuracy of 0.03 magnitudes.

\subsection{Radio}
15-GHz radio monitoring took place at the Ryle Telescope of the Mullard Radio Astronomy Observatory, Cambridge. The radio data are presented here as 10-minute integrations with a typical rms uncertainty of 1 mJy; the phase calibrator used was B2005+403. Further details may be found in Pooley \& Fender (1997).

We have also made use of radio data from the Green Bank Interferometer (details of which may be found in Waltman et al. 1994) which monitors this and many other sources at 2.25 and 8.3 GHz.

\section{Results}

Long-term lightcurves for BATSE, ASM, $UBVK$ band photometry and the radio are shown in Fig. 1.

Perhaps the most notable feature is the very distinct anticorrelation between the two X-ray bands during the period MJD 50200--50330. As mentioned previously, this is typical of the transition to the high/soft state and back to the hard again. Unfortunately we only have one radio point during this period -- on MJD 50226 the flux reached 18 mJy. However, this was during the transition and VLA monitoring towards the end of the high/soft state period showed that, as found during previous state changes, the radio flux had dropped to $\sim$5 mJy (Zhang et al. 1997b). Some authors (Belloni et al. 1996, Esin et al. 1998) have concluded that Cyg X-1 did not fully reach the soft/high state on this occasion in which case we may expect the radio source to not have disappeared as before. However, it should be noted that the radio source of GX339-4 diminishes during its X-ray intermediate/high state (Fender et al. 1999 in prep) and, given the many similarities between these two systems in terms of their X-ray and radio properties, we may expect Cyg X-1 to behave likewise.

It is also immediately apparent that there is a long term modulation in the radio,  referred to as the `150-day' period in Pooley, Fender \& Brocksopp (1999); it has now been monitored for five cycles. This is studied in more detail in section 3.2.

There are many flares in the X-ray and radio light curves, the majority of which seem to correlate between the different wavelength regimes. This is perhaps to be expected as the emission is all thought to be associated with the black hole and accretion disc. The majority of the X-ray flares appear to be of comparable amplitudes but there are two in the ASM data which increase above this level. The flares at MJD $\sim$50650 and $\sim51010$ are studied in further detail in section 3.2.

In contrast to this, the optical photometry shows a very flat lightcurve with few deviations, other than orbital modulation. There are a few flares but whilst the behaviour of one optical band is generally correlated with the others, there appears to be no correlation with the other wavelength regimes. In particular, there is one flare of note that occurred at MJD 50583 in the $U$ and $B$ bands and is distinctly not correlated with any soft X-ray behaviour. This is addressed in more detail in Bochkarev \& Lyuty (1998). Therefore it is perhaps surprising that the equivalent width of the H$\alpha$ emission line was apparently anti-correlated with the soft X-rays during the soft X-ray state, being significantly below its usual value -- indeed, the emission component of the line essentially disappeared, with a value of $\sim$0.25 \r{A} compared with an average value of -0.7 \r{A} during the low/hard state. We suggest that this was due to either additional ionisation in the stellar wind at this time, or dilution by a stronger free-free component. We do not include the equivalent width data in Fig. 1 as it does not cover the full period. Instead, we address it in further detail in section 3.1 and 3.3.

Unfortunately our infrared data are more scattered and it is difficult to see if there are any flares. It appears that there are none and that generally it follows the same flat shape of the optical. However, there are a number of possible deviations from this. It appears that the flux is slightly increased during the X-ray high/soft state and then falls to below the mean shortly after the transition to the low/hard state. This is the case in all four bands, although we show only the $K$ band here. We note that there is a single point in the $U$, $B$ and $V$ bands soon after this that is also well below the mean flux level. There may also be a peak in the infrared at MJD 51055, approximately forty days after the X-ray flare.

\begin{figure*}
  \begin{center}
    \leavevmode
    \vspace{1cm} 
    \psfig{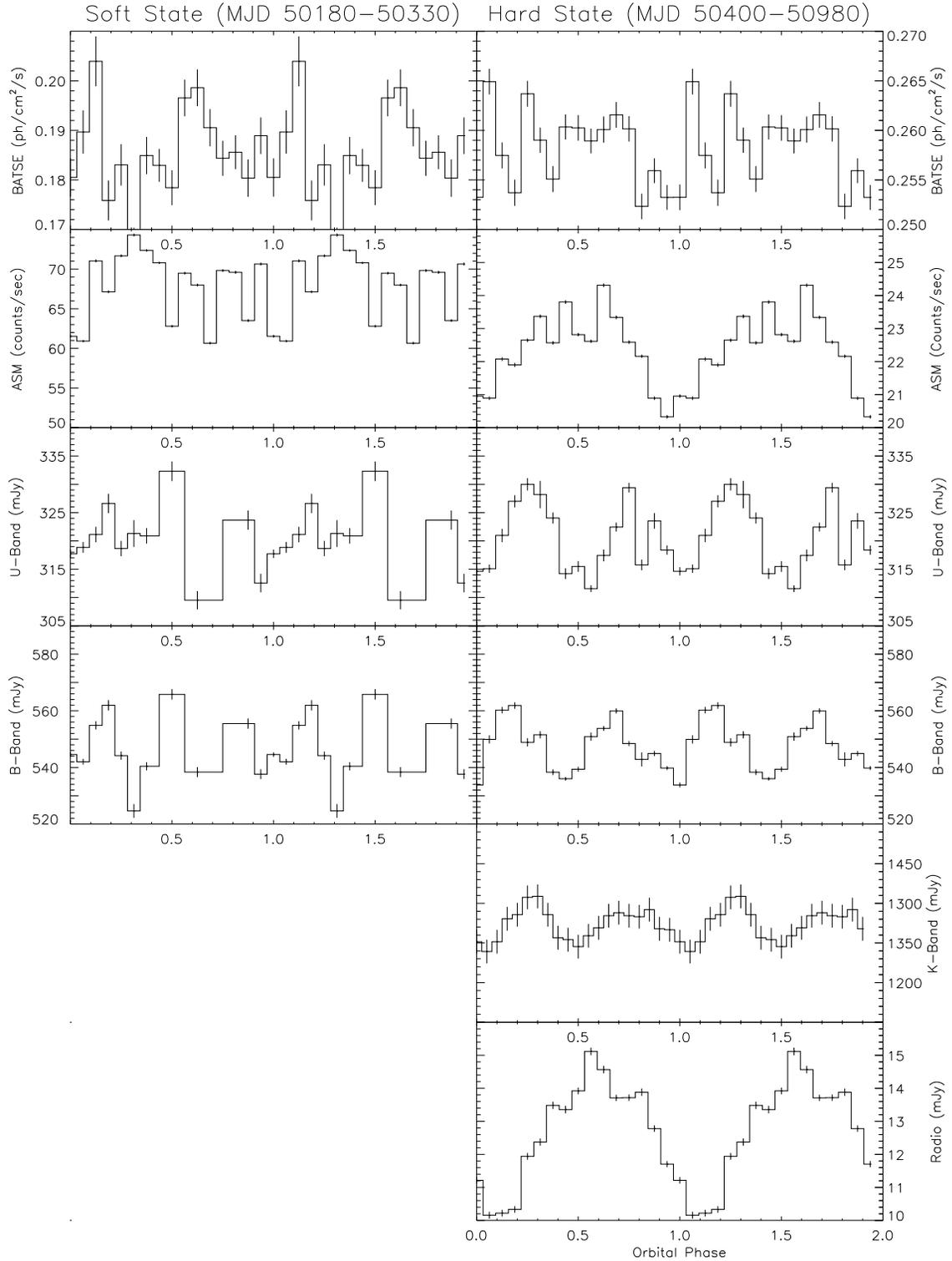} 
    \vspace{0.5cm}   
    \caption{Lightcurves folded onto the 5.6-day orbital period. Plots on the left correspond to the soft X-ray outburst of May-August 1996. Plots on the right correspond to the return to the soft X-ray low state; flares have been removed prior to folding. BATSE is at the top, then ASM, {\it U, B, K} and radio. (We only have sufficient $K$-band and radio monitoring during the hard state)}
    \label{fig:Mean Lightcurves2}
  \end{center}
\end{figure*}

\subsection{The orbital period}

We use the ephemeris of Brocksopp et al. (1999) with P$_{orb}$=5.599829$\pm$0.000016 days. The {\sc starlink} package {\sc period} was used to fold all data on the orbital period with superior conjunction of the black hole corresponding to the zero point (JD 2441874.707).

We were surprised to see that the optical photometry did not fold onto this period satisfactorily, in contradiction to Brocksopp et al. (1999) when a much longer dataset was used. We therefore separated the data according to the X-ray state and, for the hard state, removed flares and scatter above $\sim3\sigma$ from all datasets. The resulting mean lightcurves are shown in Fig. 2.

\begin{figure}
  \begin{center}
    \leavevmode
    \psfig{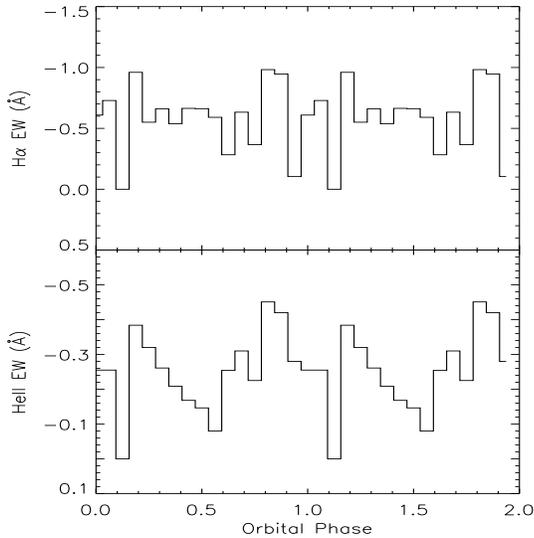} 
    \caption{All Equivalent width data of H$\alpha$ and He\,{\sc ii} folded on the orbital period. Errors $\sim10\%$.}
    \label{fig:EW}
  \end{center}
\end{figure}

\subsubsection{The low/hard state}

We now find that, during the low/hard state, the orbital modulation is much more apparent in the optical (as found by Voloshina, Lyuty \& Tarasov 1997) and the soft X-ray (this was also found by Cui, Chen \& Zhang 1998). Contrary to this, when the flares/scatter (above $\sim3\sigma$) have been removed, the BATSE data no longer appears to modulate on the orbital period, despite clearly showing modulated behaviour in Paciesas et al. (1997). This suggests that either the flares occur periodically or some component modulates during the flaring periods. The radio modulation does not change significantly when the flares are removed.

The optical and infrared show the double-peaked ellipsoidal modulation that we would expect from a star with a tidally distorted companion. As we have insufficient infrared data from the time of the 1996 high/soft state we cannot say whether or not the orbital modulation is disrupted. The amplitude of the infrared modulation is comparable with that of Nadzhip et al. (1996) ($\sim4\%$), not with that of Leahy \& Ananth (1992) ($\sim7\%$). 

Pooley, Fender \& Brocksopp (1999) discovered that the 15-GHz radio phase is offset from the ASM data by about 0.67 days. This delay was also apparent at other radio frequencies and increased with wavelength. We find that extending the dataset does not dispute this result and also notice that there is a phase difference between the minima of {\em all} our datasets (R.H.S. of Fig. 2) -- the optical reaches its minimum at phase zero (by definition) with the X-rays slightly beforehand and the infrared and radio at increasing longer delays. It is not clear exactly where these lags originate but this is probably an optical depth effect. It is interesting that there is a similar phase delay between the hard and soft X-rays in Cyg X-3 (Matz 1997).

We also fold all of our equivalent width data for H$\alpha$ and He\,{\sc ii}\,$\lambda 4686$ (Fig. 3); errors on each point are $\sim10\%$. Fitting a double peaked sine wave to the two datasets yields a reduced $\chi^2 \sim$ 0.07 for H$\alpha$ and 0.01 for He\,{\sc ii}, suggesting that the apparent structure seen in at least the latter may be real. Further analysis of our spectra will occur in a later paper.

\subsection{The long-term period}

\begin{figure}
  \begin{center}
    \leavevmode
    \psfig{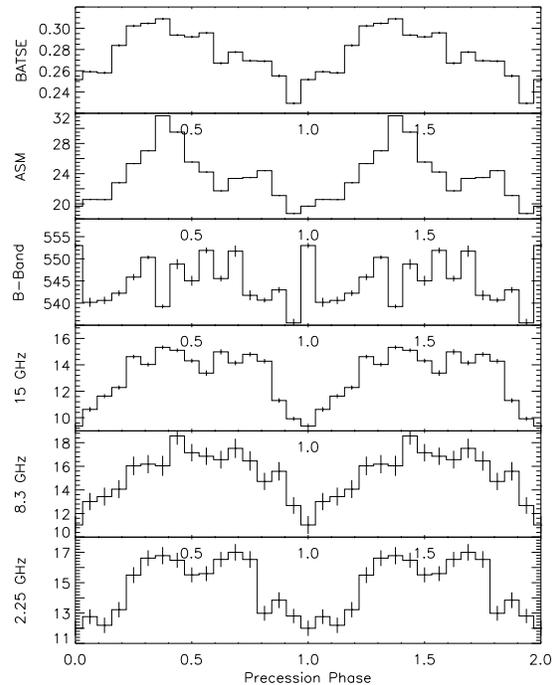} 
    \caption{X-ray, $B$ band photometry and radio (including GBI at 2.25 and 8.3 GHz) data folded on the 142-day period (low/hard state only). All units are mJy except for BATSE (photon/cm$^{2}$/s) and ASM (counts/s)}
    \label{fig:Long Period}
  \end{center}
\end{figure}

A Lomb--Scargle period search suggests that there is also a period of 142.0 $\pm$ 7.1 days present in the radio data (the `150-day period' of Pooley, Fender \& Brocksopp 1999), although it is clearly not sinusoidal (Fig. 4). We also include 2.25 and 8.3 GHz radio data from the Green Bank Interferometer which shows that this time the amplitude of modulation and the time of phase zero do not show the strong wavelength-dependence exhibited by the orbital modulation. $T_0$ corresponds to JD2450395.19$\pm$3. This period can also be detected in the X-rays and optical; the $B$-band is included in Fig. 4 and, while it does not initially appear to show such definite structure, it is only an excess at phase 0.0 that makes this modulation dubious and we find that this excess is due to the flare at $\sim$ MJD50670. Perhaps surprisingly, we find no evidence for the 294 day period found by Priedhorsky, Terrell \& Holt (1983) and Kemp et al. (1983, 1987).

(N.B. Fig. 4 includes data from the hard X-ray state only. We also note that the X-ray flux in Fig. 4 appears different from that of Fig. 2 (R.H.S.). This is because flares have {\em not} been removed before folding the data on the 142-day period.)

Bruevich et al. (1978) calculate the disc contribution of the optical luminosity to be $\sim2$\% which is equivalent to the apparent modulation of the B Band in Fig. 4. We can therefore assume that the 142-day period in the optical is related to the accretion disc, rather than the supergiant.

If we are to assume that the modulation is due to the precession of the accretion disc (as found in other X-ray binaries) then there is no reason to expect the period to remain constant over a decade. This is particularly true for a system accreting via a wind, which has a more variable accretion rate than for Roche lobe overflow. To investigate this further, we have looked at the public access BATSE data from before the 1996 high/state transition. While before this time there was some modulation at 142 days, it was considerably less than since the state change. Incidentally, the 294-day modulation was slightly stronger before the transition but is not present in the last 2.5 years' worth of data. We therefore assume that it is indeed possible for the precession period to change.

However, Larwood (1998) shows that the ratios of the orbital and precessional periods for other X-ray binaries with precessing discs are approximately equal; 294 days is too large for Cyg X-1 to show a similar ratio, whereas 142 days is much more suitable. We use the calculations of Larwood (1998) to calculate the angle, $\delta$ between the inclined disc and the orbital plane.

\begin{equation}
\frac{P}{P_{p}}=\frac{3}{7}\beta^{3/2}\frac{\mu R^{3/2}\cos\delta}{(1+\mu)^{1/2}}
\end{equation}

\hspace*{-0.5cm} where $P$ is the orbital period, $P_p$ is the precessional period, $\mu$ is the mass ratio, $R$ is the ratio between the Roche radius and the accretion disc radius and $\beta$=$\beta_{P}$=Paczy\'{n}ski's radius (Larwood 1998, Paczy\'{n}ski 1977). $R$ and $\beta_P$ are given by:

\begin{equation}
R(\mu)=\frac{0.49}{0.6+\mu^{3/2}\ln(1+\mu^{-1/3})}
\end{equation}

\begin{equation}
\beta_{P}(\mu)=\frac{1.4}{1+[\ln(1.8\mu)]^{0.24}}
\end{equation}

This gives an angle $\delta$=37$^{\circ}$ between the disc and the orbital plane.

For an accretion disc of area $A$ inclined at angle $i$, the orbital inclination, the projected area observed is $A\mbox{cos}i$. When the disc is tilted towards us through angle $\delta$ the projected area is increased to $A\mbox{cos}(i-\delta)$ and the soft X-ray emission will be a maximum. Likewise the soft X-ray emission will be a minimum for a projected area of $A\mbox{cos}(i+\delta)$.

From Fig. 4 the ratio of soft X-ray maximum to minimum emission is $\sim1.62$. We use this to deduce that Larwood's model fits our data if we assume $i=20^{\circ}$. This is rather low (e.g. Bruevich et al. (1978) suggest $i=40-60^{\circ}$, Kemp et al. (1979) $i=60-70^{\circ}$) and so it appears that some refinement to the model is necessary.

The radiative warping model of Wijers \& Pringle (1998) predicts a long-period of 180 days for Cyg X-1. Again, our value of 142 is closer than 294 but it still suggests that some aspect of the model does not fit. This model invokes more parameters which are not well-known, such as viscosity and mass accretion rate rather than just the mass ratio, and so it is not surprisingly that the predicted and actual values do not agree. 

\begin{figure}
  \begin{center}
    \leavevmode
    \psfig{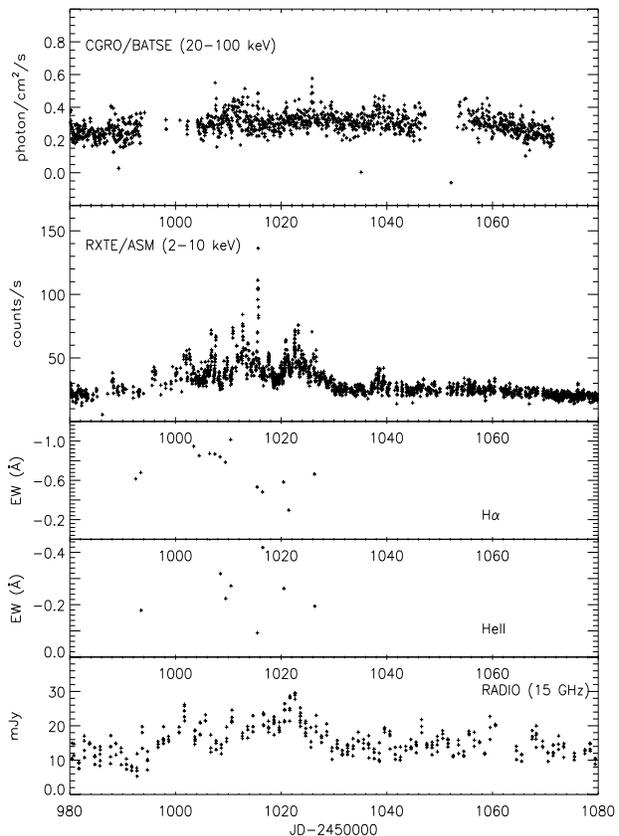} 
    \vspace{0.5cm}   
    \caption{X-ray and radio lightcurves and equivalent widths during flare period MJD 50980--51080}
    \label{fig:Flares}
  \end{center}
  \vspace{-0.25cm}
\end{figure}

\subsection{Correlated Flares}

We now consider the two larger flares of MJD 50630--50670 and MJD 50980--51080 in more detail. 

During the first of these periods it can be seen that there is a pair of X-ray flares (see Pooley, Fender \& Brocksopp 1999 for a more detailed plot), simultaneous in both energy bands with the radio `flare' occurring in between. (On removal of the orbital period this radio flare is more pronounced). The second flaring period (shown in Fig. 5) is very different as this time there does not seem to be any obvious correlation between ASM and BATSE. It is clear from Fig. 1 that the hard X-ray lightcurve changes very little compared with the ASM and radio, despite some violent activity in the accretion disc. We also note that the radio is much more active than previously -- rather than just a small flux increase it appears to roughly follow the shape of the ASM lightcurve.

\begin{table}
\caption{Spearman rank correlation coefficients ($r$) of plots in Fig. 6}
\label{correlation coefficients}
\begin{center}
\begin{tabular}{lcccc}
\hline
&Hard State&Flare 1&Flare 2&Soft State\\
\hline
ASM/BATSE&0.59&0.83&0.39&$-$0.54\\
ASM/Radio&0.30&0.30&0.70&--\\
BATSE/Radio&0.25&0.27&0.31&--\\
\hline
\end{tabular}
\end{center}
\vspace*{-.5cm}
\end{table}

\begin{figure*}
  \begin{center}
    \leavevmode
    \vspace{.5cm} 
    \psfig{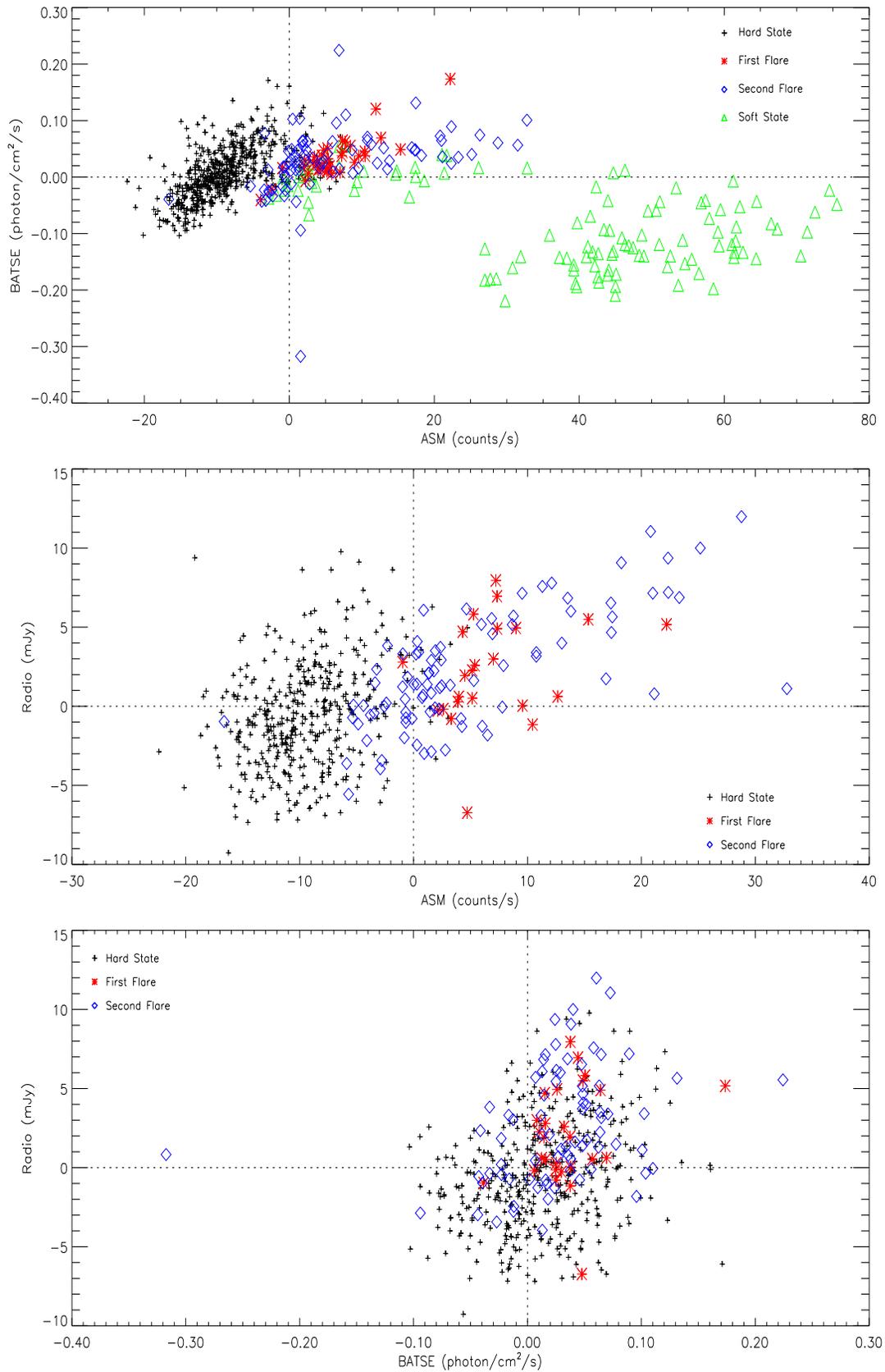} 
    \vspace{-0.25cm}   
    \caption{X-ray and radio flux-flux plots.}
    \label{fig:Flux-Flux Plots}
  \end{center}
\end{figure*}

Given the strong anti-correlation between ASM and BATSE during a state change and the strong correlation during the first flare, it is possible that in this second flare we are seeing some sort of hybrid between the two. This may have been observed before  -- Cui, Chen \& Zhang (1998) comment on a flare just prior to the state change during which the ASM spectral behaviour was similar to that of the high/soft state, but clearly not a state change. Maybe this flare is also a precursor to a state change, or possibly a `failed' state change accompanied by a radio flare which may be a typical early signature of a transition to the high/soft state (Zhang et al. 1997b).

The equivalent widths do not really show any behaviour related to the accretion disc activity. Both flare periods show a decay in equivalent width for H$\alpha$ but as single system supergiants also show a similar equivalent width variability it is presumably independent of the X-ray behaviour.

We then generate flux-flux plots for the X-rays and radio. (It is clear from Fig. 1 that to plot radio or X-rays against the photometry would yield nothing. However, flux-flux plots for the optical only show very strong correlations confirming that all bands share a common emission region). To do this we have removed the mean orbital light curves (un-normalized), firstly because we know each data set shows orbital modulation and it is other correlations that we wish to find and secondly because the radio is offset from the X-rays by 0.12 in orbital phase. We note that by removing the mean lightcurve some of the residuals become negative -- this is because we have not removed flaring periods prior to subtracting the mean lightcurves. The fluxes were then binned on 1 day and the fluxes of corresponding bins plotted.

Fig. 6 shows the three flux-flux plots. Stars and hollow shapes correspond to the different flare/transition periods and the intervening low/hard state in between is shown by crosses.

It is immediately clear that there are a number of correlations. We have already mentioned the anti-correlation between ASM and BATSE when Cyg X-1 is in the high/soft state; this can be seen in the top plot of Fig. 6. There is a large amount of scatter due to the variable flaring in the soft X-rays. During the quiescent low/hard state there is a reasonable correlation between the two X-ray bands, although the radio only seems loosely correlated with the X-rays; the possible offset between the radio and X-ray long-term period seen in Fig. 4 may be adding to the scatter. There is a strong correlation between the two X-ray datasets during the first flare, but again, only scatter in the radio/X-ray plots. Finally, a strong correlation between the radio and soft X-rays can be seen during the second flare.

\begin{table*}
\vspace*{.25cm}
\centering
\caption{Summary of the orbital and precessional modulations across the spectrum and their physical interpretations. The modulation is calculated by (max-min)/(max+min)$\times100$. For BATSE we use the dataset prior to removal of $>3\sigma$ points} 
\vspace*{1cm}
\begin{tabular}{ccccc}
\hline
     & \% Orbital modulation& Physical & \% Precessional modulation& Physical\\
     & (Standard error $\times10$)& interpretation &(Standard error $\times10^3$)       & Interpretation \\
\hline
BATSE 20-100 keV & 1.9 (1.0) & Stellar wind absorption& 18.0 (0.1) & Precession of disc/jet\\
XTE 2-10 keV & 9.0 (0.3) & Stellar wind absorption & 25.8 (0.2)& Precession of disc\\
\hline
U (0.36$\mu$m) & 2.9 (1.4) & Ellipsoidal modulation & 2.6 (15.1)& \\
B (0.43$\mu$m) & 2.6 (0.7)& Ellipsoidal modulation & 1.5 (2.6)& Precession of disc ?\\
V (0.55$\mu$m) & 2.4 (0.7)& Ellipsoidal modulation & 2.1 (7.0)&\\
J (1.25$\mu$m) & 4.2 (0.3)& Ellipsoidal modulation & ? &\\
H (1.65$\mu$m) & 5.2 (0.4)& Ellipsoidal modulation & ? &\\
K (2.20$\mu$m) & 2.5 (0.2)& Ellipsoidal modulation & ? &\\
\hline
RT (15 GHz) & 19.4 (1.2)& Stellar wind absorption & 23.9 (8.3)  & Beaming/projection of jet \\
GBI (8.3 GHz) & 17.6 (0.1)& Stellar wind absorption & 25.7 (41.0) & Beaming/projection of jet \\
GBI (2.3 GHz) & 6.8 (13.4)& Stellar wind absorption & 17.2 (40.0) & Beaming/projection of jet \\  
\hline
\end{tabular}
\vspace*{.25cm}
\end{table*}

Spearman rank correlation coefficients ($r$) have been calculated for each of the sub-plots of Fig. 6 and are shown in Table 2. $r$ is defined as the ratio of the covariance of the two datasets to the product of their standard deviations and takes values in the range -1.0--1.0. Highest values are obtained for hard and soft X-rays during the first flare, high/soft state and quiescent low/hard state and for the soft X-rays and radio during the second flare. We note that a considerably higher value of $r$ can be obtained if we assume that the radio lags the X-rays by 2.5 days, during the first flare only -- a value obtained by trial and error. If we are to assume that this lag is real then it would imply that the 15-GHz radio emission becomes optically thin a maximum distance $6.5\times10^{13}$ metres from the X-ray source. However as the only lag apparent in our data is 0.67 days between the radio and X-ray orbital modulation, 2.5 days seems unlikely to be real. $r$ is not improved if we assume this shorter lag, although as the orbital period has been removed prior to creating the flux-flux plots this may be expected.

\section{Discussion}

In summary, we find that the orbital period and also a longer period modulation, probably due to precession of an accretion disc, are present at all wavelengths. We also find that there is correlated behaviour between the radio and X-rays although the nature of this correlation varies from flare to flare. This implies that the X-rays and radio are coupled as expected for a disc/jet system. The optical and infrared do not show any behaviour related to the X-rays or radio and appear to share a separate emitting region.

It is necessary to draw together all of the results obtained for Cyg X-1 and create a consistent model that includes all available data --  not just, for example, the X-rays as in the case of some previous works. This model needs to address:

\begin{itemize}
\item The orbital modulation across the spectrum during the low/hard state
\item The long-period (precessional) modulation across the spectrum during the low/hard state
\item The lack of optical and soft X-ray orbital modulation during the high/soft state
\item Decrease of radio emission during high/soft state
\item Existence of correlated flaring behaviour between radio and X-rays
\item The occasional lack of correlation between hard and soft X-rays
\item The general lack of non-modulated variability in the optical and infrared lightcurves, other than occasional flares which appear unrelated to X-ray behaviour
\end{itemize}

The details of the periodic behaviour of Cyg X-1 is summarised in Table 3. The percentage modulations are calculated using the formula $$\frac{(max-min)}{(max+min)}\times100$$ 

The orbital modulation of the optical and infrared is due to the ellipsoidal shape of the supergiant, both due to distortion of the star itself and due to the shape of the focussed stellar wind. In the case of the radio and X-rays, the frequency-dependent phase lags and degree of modulation suggests line-of-sight absorption by the wind to be responsible. Stellar wind absorption could not account for the long-period variability, particularly as the degree of modulation does not vary with frequency, and so precession and/or radiative warping of the accretion disc seems the most likely cause. This is supported by our calculations in section 3.2.

To account for both periods we adopt the conical jet model of Hjellming \& Johnston (1988) for the radio emission (detailed radio modelling will be presented in a later paper). The production of continuous synchrotron emission requires a continuous flow of relativistic electrons and magnetic field from the accretion disc. We therefore find it extremely unlikely that this could occur throughout the proposed corona that is currently the only accepted model for the X-rays. A collimated jet is much more feasible and we see no reason why the hard X-rays are not emitted from the base of this jet, produced when seed photons from the accretion disc are up-scattered as previously suggested. The jet model is also supported by VLBA observations of Stirling et al. (1998) and de la Force et al. (1999, in prep).

\begin{figure*}
  \begin{center}
    \leavevmode
    \vspace{1cm} 
    \psfig{file=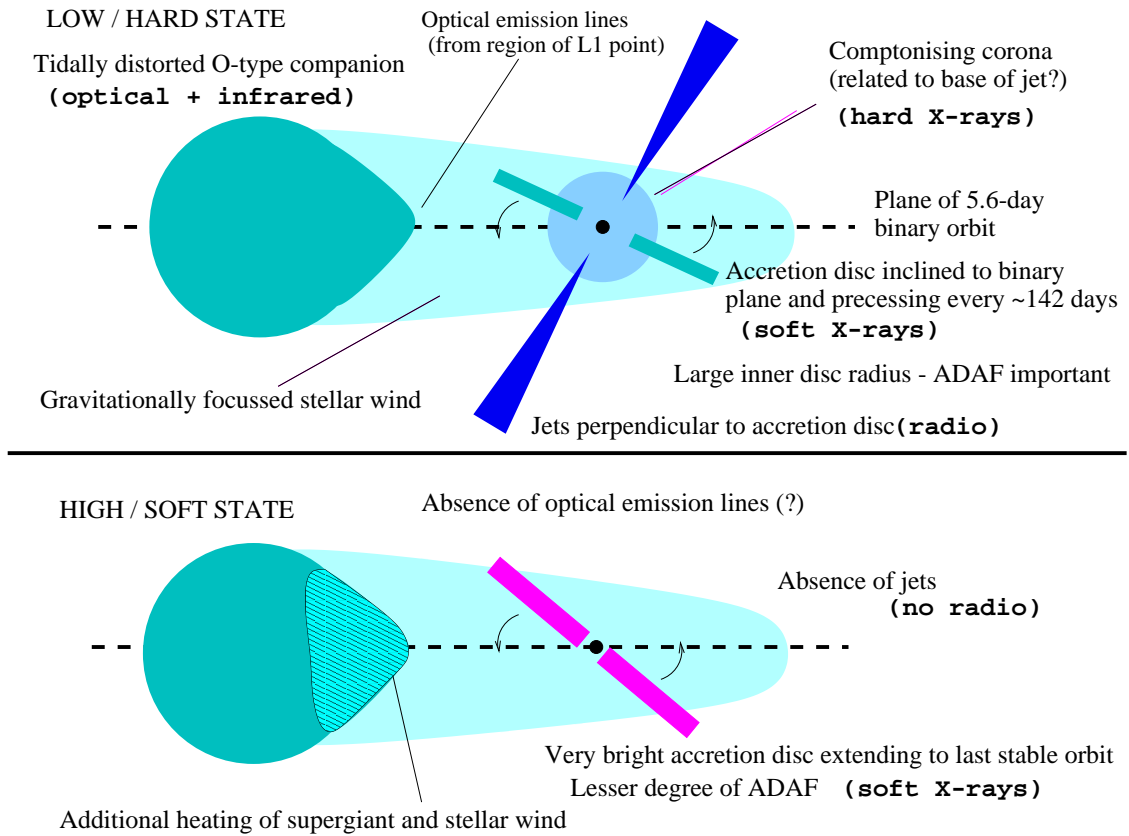, angle=0, width=15cm} 
    \vspace{0.25cm}   
    \caption{Schematic showing the emission regions of Cyg X-1 during the hard and soft X-ray states.}
    \label{fig:Cyg X-1}
  \end{center}
\end{figure*}

The apparent precession of the X-rays questions previous models of Cyg X-1. While an optically thick corona could precess with the accretion disc providing it had some non-uniform geometry, it is generally accepted that the corona is optically thin (e.g. Cui et al. 1997). In this case the corona would not precess with the disc unless it was a very thin layer {\em just} above and below the disc, and therefore the BATSE modulation should be comparable with that of ASM. This is not so and it can be seen in Table 3 that the BATSE modulation is considerably less than that of ASM; in fact it is more comparable with that of the radio. This suggests that the jet may be responsible for the production of the hard X-rays. Likewise if, as predicted, the low/hard state is explained by an optically thin advection dominated accretion flow (ADAF) then neither should the soft X-rays display the precession modulation. Therefore, {\em either} the soft X-ray flux during the low/hard state does not entirely come from an optically thin ADAF {\em or} we must assume a radiation-driven precession model. Radiation warping could cause the disc to become flared and so partial eclipses would produce the modulation. However, we feel that the Wijers \& Pringle (1998) model is not able to fit our data quantitatively and some further refinements are necessary.

The increased mass flow during the high/soft state results in soft X-ray flaring of a magnitude much greater than the amplitude due to the orbital variability which is why this modulation appears to disappear. The lack of orbital modulation in the optical at this time is not so easily explained. Voloshina, Lyuty \& Tarasov (1997) discovered that the orbital modulation does not `disappear' but becomes single peaked with a maximum occurring at the same phase as the X-rays, rather that the standard double peaked ellipsoidal variability. This might suggest that, during this period, the luminosity of the accretion disc was higher than the supergiant. However, with a disc contribution of only $\sim2$\% (Bruevich et al. 1978), this is highly unlikely and we think that either the accretion disc is bright enough for additional heating or ionization of the supergiant to become significant and observable at orbital phase 0.5 or we are seeing the accretion stream from the L1 point.

Our results support the theories in which free-free emission in the disc produces the soft X-ray emission. Some of these X-rays (and perhaps also lower energy photons in the disc) are then up-scattered by a hot corona either side of the disc, via the inverse Compton mechanism, to higher energies. Acceleration of disc electrons along magnetic field lines gives rise to synchrotron radio emission. With such constant radio flux levels, a continuous injection of electrons is required and so the radio emission must be coming from collimated jets, rather than a more diffuse corona. It is therefore possible that the hard X-ray flux comes from a smaller region near the base of the jet than thought previously. We also use the equations of Marscher (1983) to calculate that the contribution to the hard X-ray flux of synchrotron self-Compton scattering by the jet is probably negligible ($<<1$\%).

The different types of correlated and anti-correlated behaviour can be explained in terms of different events taking place within the system. It appears that small flares occurring in the X-rays and radio are a result of slightly increased mass flow. Transitions to the high/soft state also involve increased mass flow but, in addition, the inner disc radius decreases considerably and a lesser degree of advection takes place.  Where there appears to be no correlation or anti-correlation between the hard and soft X-rays it is possible that we have a `failed' state change; indeed, we find spectral softening in both ASM and BATSE data at the time of this second flare. Although the radio states are generally thought to be anti-correlated with the soft X-rays, it seems that the {\em transitions} from one state to another are accompanied by radio flares which may explain the relatively strong radio flux at this time.

The apparent lack of correlation between the photometry and the X-rays during the low/hard state is not surprising as the supergiant is so bright that any additional luminosity due to irradiation is negligible. Indeed, we show in Brocksopp et al. (1999) and Lyuty (1999 in prep.) that the small deviations from orbital modulation are a result of the wind, although we have shown in section 3.2 that there is also a small contribution from the accretion disc.

In drawing everything together we find that previous models need slight alterations to explain all of our results. Fig. 7 shows a schematic of Cyg X-1 which tries to include all information obtained about the system.

During the low/hard state optical and infrared emission is dominated by that of the stellar wind of the supergiant. In particular, the two emission lines H$\alpha$ and He\,{\sc ii}\,$\lambda 4686$ are produced in the accretion stream, near the L1 point. At this time, the soft X-ray emitting accretion disc has a large inner radius and advection dominated flow. It is inclined to the orbital plane of the system and precesses with a period of $\sim$142 days by means of a radiation-driven mechanism. Small radio jets are emitted from the centre of the disc and precess with it. Hard X-rays are emitted as the result of upscattering of the soft X-ray photons to higher energies by a small corona at the base of the jet. 

When Cyg X-1 enters the less common soft X-ray state the accretion disc extends much further towards the black hole and the ADAF is restricted to the corona (Esin et al. 1998). Increased activity in the disc causes additional heating and ionization of the stellar wind, resulting in a loss of the H$\alpha$ emission line. The jets and corona disappear, presumably disrupted by either the inner edge of the accretion disc or the heated stellar wind and the material advected into the black hole.

\section*{acknowledgements}
This work was completed on the Sussex {\sc STARLINK} node. We thank Michiel van der Klis and Mariano Mendez for their comments about the accretion disc corona and Guillaume Dubus for a useful discussion about disc precession. We are very grateful for the quick-look results provided by the ASM/RXTE team and to the many people who have taken observations for us. The Ryle Telescope is supported by PPARC. The Green Bank Interferometer is a facility of the National Science Foundation operated by the NRAO in support of NASA High Energy Astrophysics programs. 

CB acknowledges a PPARC studentship. RPF is an EC Marie Curie Fellow supported by grant ERBFMBICT 972436. VML is grateful to RFBR (grant 98-02-17067) for support in part and VL to RFBR (grant 96-02-19703) and the Russian Federal Program `Integration' (grant K0 232). AET and PR thank the Royal Society for a grant supporting collaboration with the former Soviet Union.


\bibliographystyle{mn}

\bsp 

\label{lastpage}

\end{document}